\documentclass[a4paper,11pt]{article}
\usepackage{pos}
\usepackage{mathtools}
\usepackage{amsmath}

\title{Transport and Connection to Heavy-ion Collisions via Heavy Flavor Probes}
\author*[a]{Hai-Tao Shu}

\affiliation[a]{Physics Department, Brookhaven National Laboratory, Upton, New York 11973, USA}

\emailAdd{hshu@bnl.gov}

\abstract{
The heavy ion experiments in Relativistic Heavy Ion Collider (RHIC) and Large Hadron Collider (LHC) are going through upgrade in the next five years, shifting their focus more on the hard processes in the new runs. One of the main goals is to draw a finer image for the quark gluon plasma (QGP). The heavy flavor probes , which witness the whole history of heavy ion collision are particularly sensitive to test the properties of QGP formed
in such collisions. The lattice results for heavy flavor probes provide transport and phenomenological models crucial inputs to describe the experimental observations like the strong suppression of the nuclear modification factor $R_{AA}$ and the non-zero azimuthal anisotropy at low $p_T$. In the last two years we have seen significant advances in the lattice QCD studies of  heavy flavor probes, including the in-medium quarkonium properties, the complex static quark-antiquark potential and the heavy quark diffusion from lattice simulations at nonzero temperature. These achievements substantially deepen our understanding of the fate of quarkonium, the screening/unscreening of the complex potential and the temperature and quark mass dependence of the heavy quark diffusion in thermal medium. In these proceedings, we review recent results and briefly discuss possible directions in these studies.

}

\FullConference{The 40th International Symposium on Lattice Field Theory (Lattice 2023)\\
July 31st - August 4th, 2023\\
Fermi National Accelerator Laboratory\\}

\begin{document}
\maketitle

\section{Introduction}

Studying the properties of QGP matter created in heavy ion collisions (HICs) \cite{Collins:1974ky,STAR:2005gfr,BRAHMS:2004adc} is complicated, for the reason that most constituents experience multiple scattering and are confined again into hadrons in a very short time, leaving us no direct access to them. 
%Analyzing the yields and momentum distributions collected at the detectors to extract the information of QGP is indirect and difficult. T
Heavy flavor probes can provide more direct access to the properties of QGP. 
Heavy flavor probes are produced in the initial stages of the collision
and at later stages of the collisions they are neither created or destroyed.
They participate in the entire dynamics of the system 
produced at  HIC, but are much less affected by the late stage hadronic
interactions 
thanks to their large mass. Thus have been intensively used both experimentally \cite{Rapp:2018qla,Dong:2019unq} and theoretically \cite{He:2022ywp} to study the properties of QGP.
The heavy flavor probes include open heavy flavor hadrons and heavy quarkonia.
Heavy quarkonia can probe the created medium at different length scales, meaning that different states, e.g. $\Upsilon$ 1S, 2S and 3S, are of different size and their dissociation follows a hierarchy pattern. By investigating the fate of different states we gain resolution to different stages of the thermalization.

\begin{figure*}
\centerline{
\includegraphics[width=0.52\textwidth]{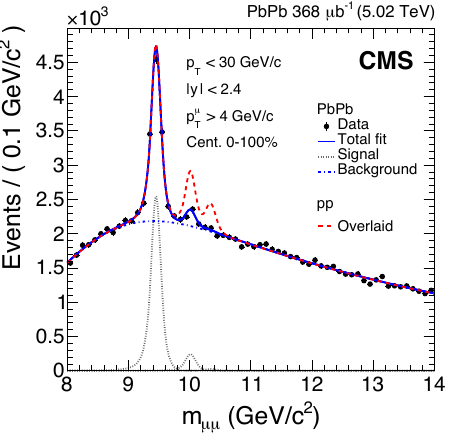}
\includegraphics[width=0.48\textwidth]{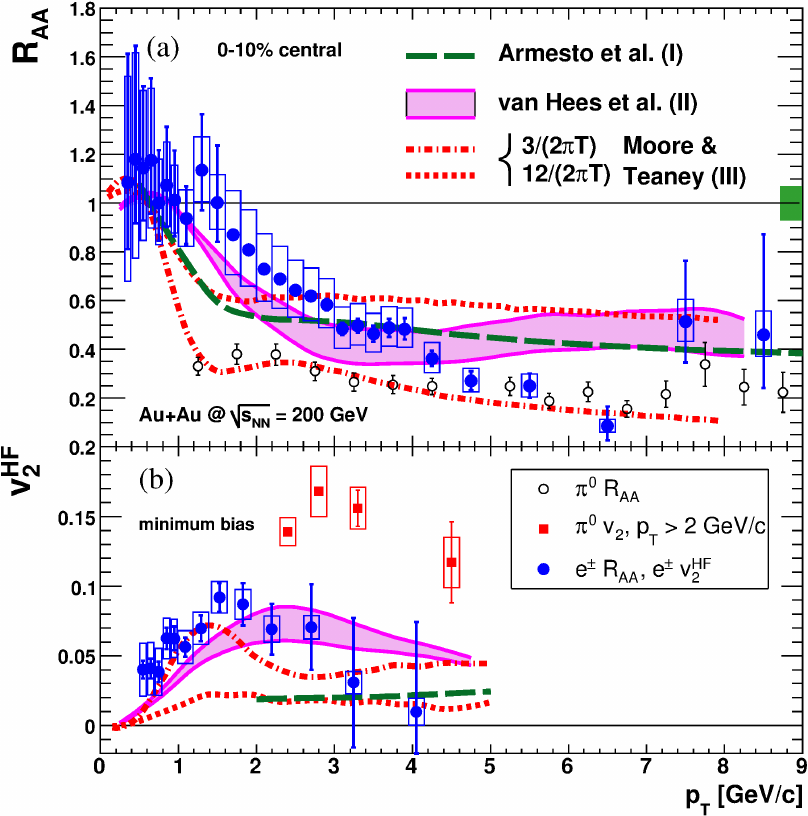}
}
\caption{Left: the sequential suppression in the production of vector bottomonium at CMS \cite{CMS:2017ycw}. Right: the suppression of the nuclear modification factor $R_{AA}$ and the non-zero azimuthal anisotropy $v_2$ at low $p_T$ observed at PHENIX \cite{PHENIX:2006mhb}.}
\label{fig:Raa-v2}
\end{figure*}

Heavy ion collisions can be produced in the large collision facilities like RHIC at BNL and LHC at CERN. After the upgrade of the heavy ion experiments the focus will shift more to the hard process \cite{Aprahamian:2015qub}. One famous phenomenon seen in the HICs is the sequential suppression in the production of heavy quarkonium, e.g., CMS Collaboration found that higher excited $\Upsilon$ (vector bottomonium) gets more suppressed in the presence of hot QGP medium \cite{CMS:2017ycw}, see left panel of Fig.\ref{fig:Raa-v2}. There is a long history of trying to understand this suppression. In 1980s Matsui and Saltz proposed the concept of color screening \cite{Matsui:1986dk} and argued that color screening  will prevent quark-antiquark pair forming a quarkonium, and this in turn will lead to the suppression of quarkonium
yields in HIC. The larger is the size of the quarkonium, 
the stronger is the effect
of color screening on its binding. This idea relies on a non-relativistic potential picture of quarkonium binding and the screening of the potential. At zero temperature the potential description turned out to be successful in describing the phenomenology of the ground state and the excited states below the open heavy flavor threshold, see e.g. \cite{Quigg:1979vr, Eichten:1995ch}. At some level it can be 
justified through the use of EFT approach \cite{Brambilla:2004jw}.
The effective theory approach leading to the potential picture can also be
generalized to the case of non-zero temperature, but the potential becomes complex \cite{Burnier:2016kqm, Laine:2006ns, Brambilla:2008cx}, and in general the real part of the potential will not be screened \cite{Brambilla:2008cx}. Lattice QCD calculations 
of the potential appeared recently \cite{Bazavov:2023dci} and we will discuss them in these proceedings. Another phenomenon is the strong suppression of the nuclear modification factor $R_{AA}$ and the non-zero azimuthal anisotropy $v_2$ at low $p_T$ observed in central Au-Au/Pb-Pb collisions relative to proton-proton collisions, see e.g. \cite{PHENIX:2006mhb, ALICE:2014qvj, CMS:2017ycw, CMS:2018zza, ALICE:2018wzm} (right panel of Fig.\ref{fig:Raa-v2}). This indicates substantial energy loss of heavy quarks when moving in the medium and that the heavy quarks participate in the collective motion of the medium \cite{ALICE:2013olq,PHENIX:2014rwj,Vertesi:2014tfa} . From these we can infer that there exists a strong coupling between heavy quarks and the medium. Modeling of these two phenomena suggests that the medium is almost a perfect fluid. Nevertheless, the figure also shows that describing both phenomena simultaneously using modeling is challenging. Such puzzle demands input information on the formation and evolution of QGP, like the shear and the bulk viscosity, in-medium dissociation temperature of quarkonium, and heavy quark diffusion coefficient. However, determination of these in perturbation theory is challenging \cite{He:2014epa,Cao:2014fna,Cao:2018ews,Li:2021xbd} as they are intrinsically non-perturbative quantities. This is feasible from the lattice calculations, though.

In these proceedings we review the recent progresses achieved in the following three topics related to  heavy flavor probes: the in-medium quarkonium properties, the inter quark potential and the heavy quark diffusion.

All the desired information mentioned above can be obtained from the spectral representation of the target objects (denoted as $X$) embedded in the corresponding Euclidean correlation function. The spectral function and correlation function are related via a Laplacian convolution equation
\begin{equation}
  G^{ }_X (\tau, T)\equiv \int \mathrm{d}\vec{x}\left\langle \mathcal{O}_X(\vec{x}, \tau) \mathcal{O}^{\dagger}_X(\vec{0}, 0) \right\rangle = \int_0^\infty
 \frac{{\mathrm{d}}\omega}{\pi} \rho^{ }_X (\omega, T) K(\omega,\tau,T),
\label{eq:corr-spec}
\end{equation}
Where  
\begin{equation}
K(\omega,\tau,T)= \frac{\cosh \left( (\frac{1}{2T} - \tau )\omega \right) }
 {\sinh\left( \frac{\omega}{2 T} \right) }
\label{eq:kenel}
\end{equation}
for most of cases. However, when considering quarkonium in non-relativistic QCD (NRQCD)
or when calculating the potential $K(\omega,\tau,T)=\exp(-\omega \tau)$, because the
relevant correlators are not periodic in $\tau$.

The extraction of the spectral function from the correlation function is ill-posed due to the limited amount of lattice data points. Various methods \cite{Jarrell:1996rrw, Ding:2017std, Burnier:2013nla, Backus:1968svk, Dudal:2013yva, Itou:2020azb, Chen:2021giw} have been proposed to tackle this problem, including the most commonly used $\chi^2$ fitting with physically-motivated models. In the case of the vector quarkonium spectral function, the transport peak at small frequencies provide information to the heavy quark diffusion and the deformation of the resonances tells the thermal effects on the bound states. In the following sections we will revisit the above equations in individual case.

\section{In-medium quarkonium properties}

\begin{figure*}
\centerline{
\includegraphics[width=0.5\textwidth]{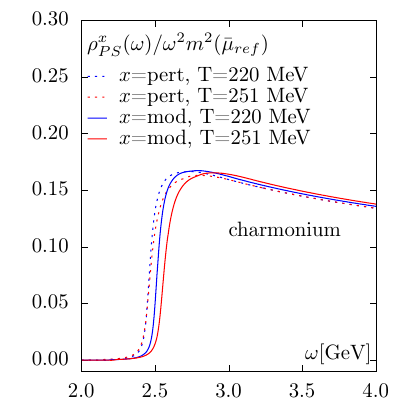}
\includegraphics[width=0.5\textwidth]{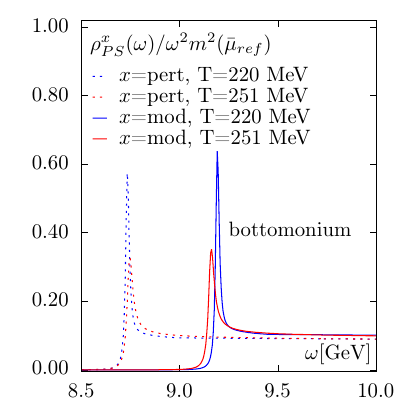}
}
\caption{The charmonium (left) and bottomonium (right) spectral function obtained in a (2+1)-flavor calculation for the pseudo-scalar channel taken from \cite{Ali:2023kmr}. The dashed curves are the perturbative models and the solid curves are the fit results.}
\label{fig:spf-cc-bb}
\end{figure*}

To study the in-medium quarkonium properties one needs to specify the operator to be the heavy quark current $\mathcal{O}=\bar{\psi}\Gamma\psi$, where different $\Gamma$ corresponds to different channel. Since the vector channel involves a transport peak, the reconstruction of the resonances in this channel is more complicated than it in the pseudo-scale channel where transport peak is absent \cite{Karsch:2003wy, Aarts:2005hg}. The dissociation temperature of the heavy quarkonium can be read off from the change of the width  of the resonances. It was intensively studied in the past decade \cite{Ding:2012sp, Ding:2017std, Kim:2018yhk, Ding:2021ise, Aarts:2011sm, Aarts:2013kaa, Aarts:2014cda}. 

Due to the large mass, heavy quark, particularly bottom quark, is difficult to accommodate relativistically on a lattice. It becomes more severe when reaching high temperatures of interest where the number of data points in temporal direction is reducing. For this reason most of such studies were trying to reconstruct quarkonium spectral function from correlators calculated using relativistic heavy quarks in the quenched approximation. 
Recently HotQCD collaboration made the first attempt going to full QCD to study the thermal effects on the in-medium charmonium and bottomonium \cite{Ali:2023kmr} via spectral analysis. In this calculation pseudo-scalar charmonium and bottomonium correlators are calculated at $T=110, 220, 251$ MeV using (2+1)-flavor  Highly Improved Staggered Quark (HISQ) action \cite{Follana:2006rc} and tree-level improved Lüscher-Weisz gauge action \cite{Luscher:1984xn,Luscher:1985zq} with sea quark mass tuned corresponding to $m_\pi\simeq 320$ MeV. With the knowledge learned from potential non-relativistic QCD (pNRQCD), a model spectral function 
\begin{equation}
\rho^\mathrm{mod}_\mathrm{PS}(\omega)=A\rho^\mathrm{pert}_\mathrm{PS}(\omega-B)
\label{eq:spfmod}
\end{equation}
can be constructed and fit to the lattice data, where $A$ and $B$ are fit parameters accounting for the normalization of the correlator and the adjustment of thermal mass shift. The results are shown in Fig.\ref{fig:spf-cc-bb}. It can be seen that for the charmonium there is no need to introduce a resonance peak down to 220 MeV while for the bottomonium the resonance peak can persist up to 251 MeV. This is consistent with a previous quenched study on large and fine lattices \cite{Ding:2021ise} suggesting that no resonance peaks for $J/\Psi$ are needed at and above 1.1 $T_c$, while for $\Upsilon$ a resonance peak is still needed up to 1.5 $T_c$. 

It is possible to increase the extent of temporal correlators for the purpose of a reliable spectral reconstruction by using anisotropic lattice, see \cite{Jakovac:2006sf} for a quenched study on quarkonium. Its extension to HISQ action just started very recently. There are also efforts trying to determine the dissociation temperature by looking at the change of the in-medium screening mass. Since screening mass is extracted from the spatial correlator, whose extent can generally be made large, such calculations can be performed in full QCD in relatively easy manner, see e.g. Refs. \cite{Karsch:2012na,Bazavov:2014cta,Petreczky:2021zmz}.

There is a possibility to avoid the difficulty of treating heavy quarks relativistically in studies where the high frequency part 
of the spectral function is irrelevant. Namely, one can use 
NRQCD, see e.g. Ref. \cite{Kolodziej:2022low}.
In NRQCD one integrates out the energy scale related to the heavy quark mass, $M_Q$, which can
be considered large compared to typical momenta inside quarkonium and the binding
energy, which are of order $\Lambda_{\mathrm{QCD}}$.
Furthermore, the heavy quark mass is much larger than the typical temperature of HIC. NRQCD effectively removes $2 M_Q$ from the lower bound of the meson spectral function but leaves the physics around bound states region unchanged. Such method reduces the complexity and is easy to implement in full QCD. There have been intensive NRQCD calculations \cite{Aarts:2014cda,Kim:2018yhk,Larsen:2019bwy,Larsen:2019zqv} over the years and now it has have been pushed to physical pion mass. 
In NRQCD the correlators are not periodic in Euclidean time, and therefore one
can access large values of $\tau$ up to $\tau=1/T$.

\begin{figure}
\centerline{
\includegraphics[width=0.5\textwidth]{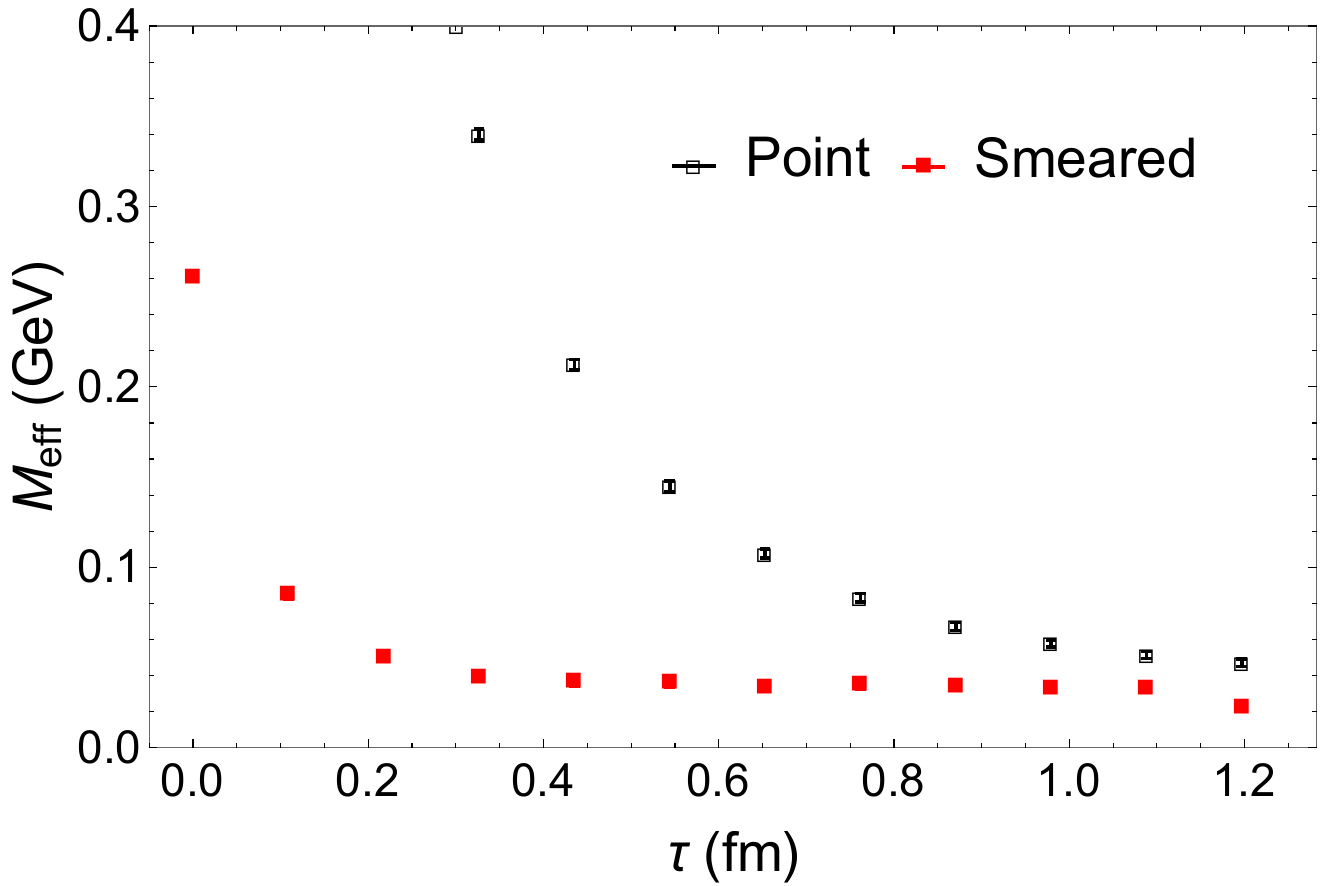}
}
\caption{Effective mass extracted from correlators calculated using point source operator and extended operator at $T<T_c$ taken from \cite{Larsen:2019bwy}.}
\label{fig:point-smear}
\end{figure}

In the most recent NRQCD calculation \cite{Larsen:2019bwy} the extended source operator technique was applied to the ground bottomonium state. It was shown that with the extended source operator we can gain better overlap with the states of desired quantum numbers, see Fig.\ref{fig:point-smear}. We can see that the effective mass reaches a plateau much earlier in the extended source case than using point source operator. This plot shows the case of low temperature. At high temperature, e.g. $T=333$ MeV that is interesting for bottomonium studies, we have access only to data at $\tau\leq 0.6$ fm for the lattice in consideration. In this case there is no hope to get the relevant physics using point sources, while 
the available time extent is sufficient when using the extended source.

\begin{figure*}
\centerline{
\includegraphics[width=0.5\textwidth]{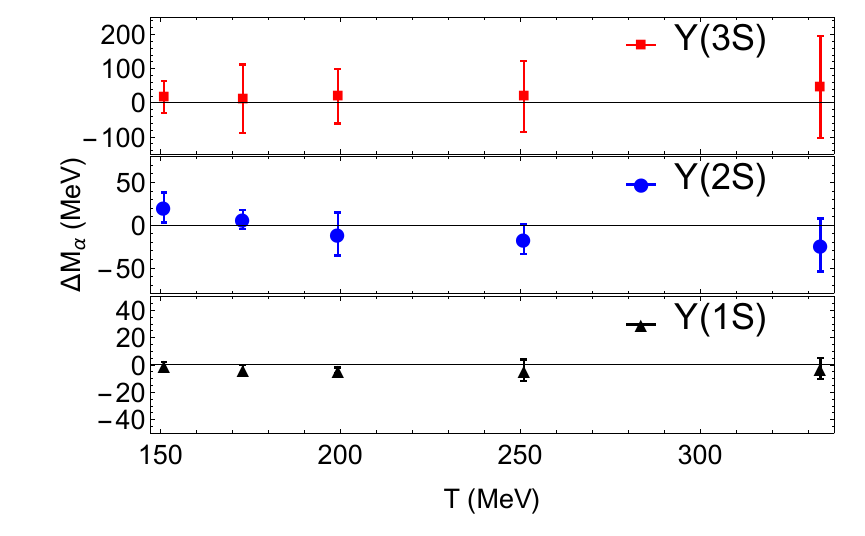}
\includegraphics[width=0.5\textwidth]{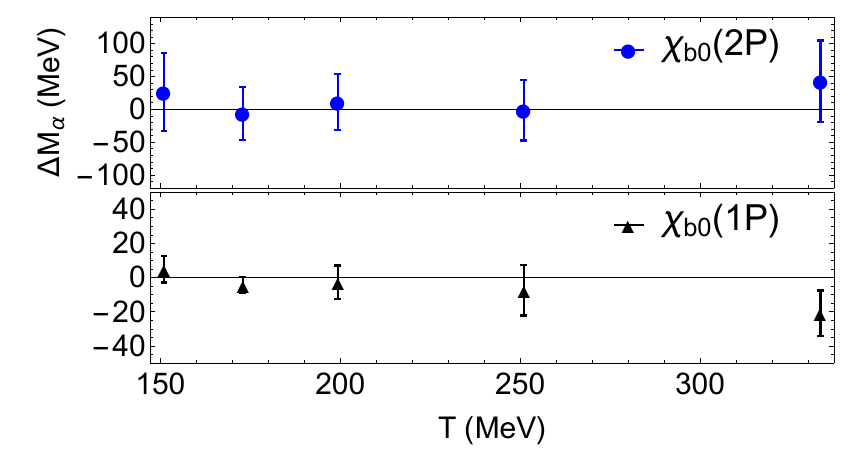}
}
\centerline{
\includegraphics[width=0.5\textwidth]{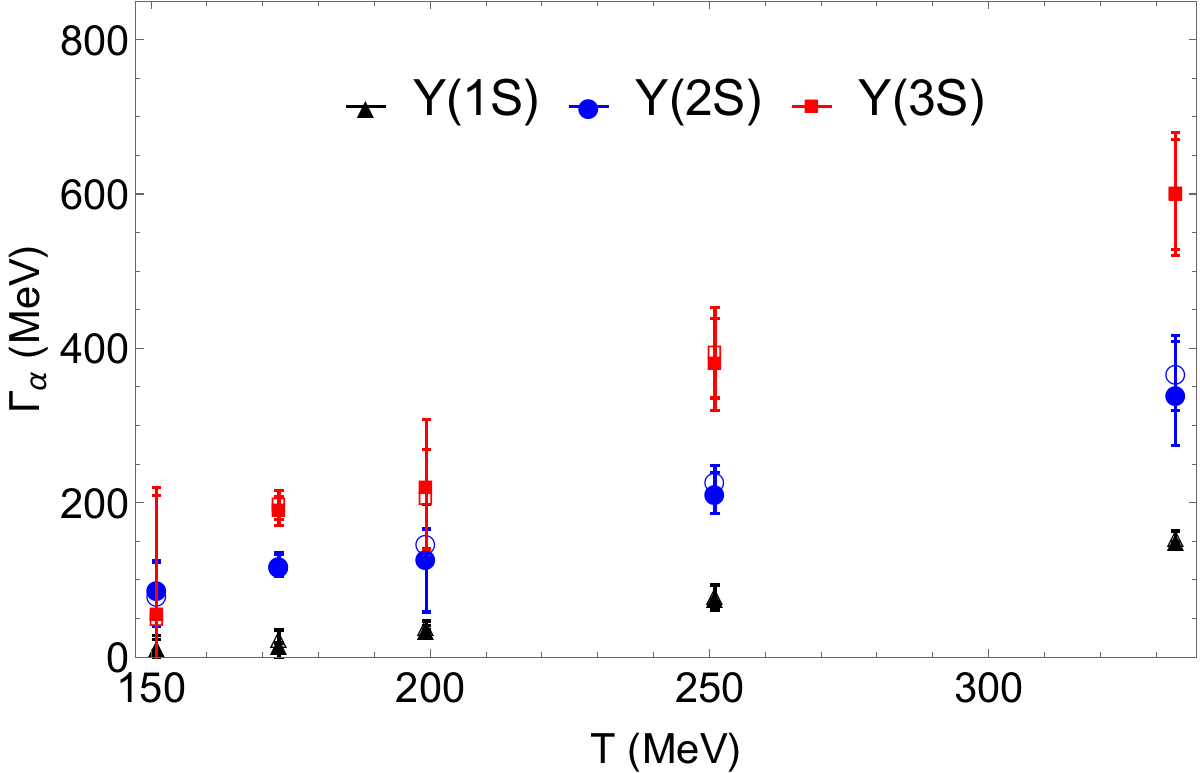}
\includegraphics[width=0.5\textwidth]{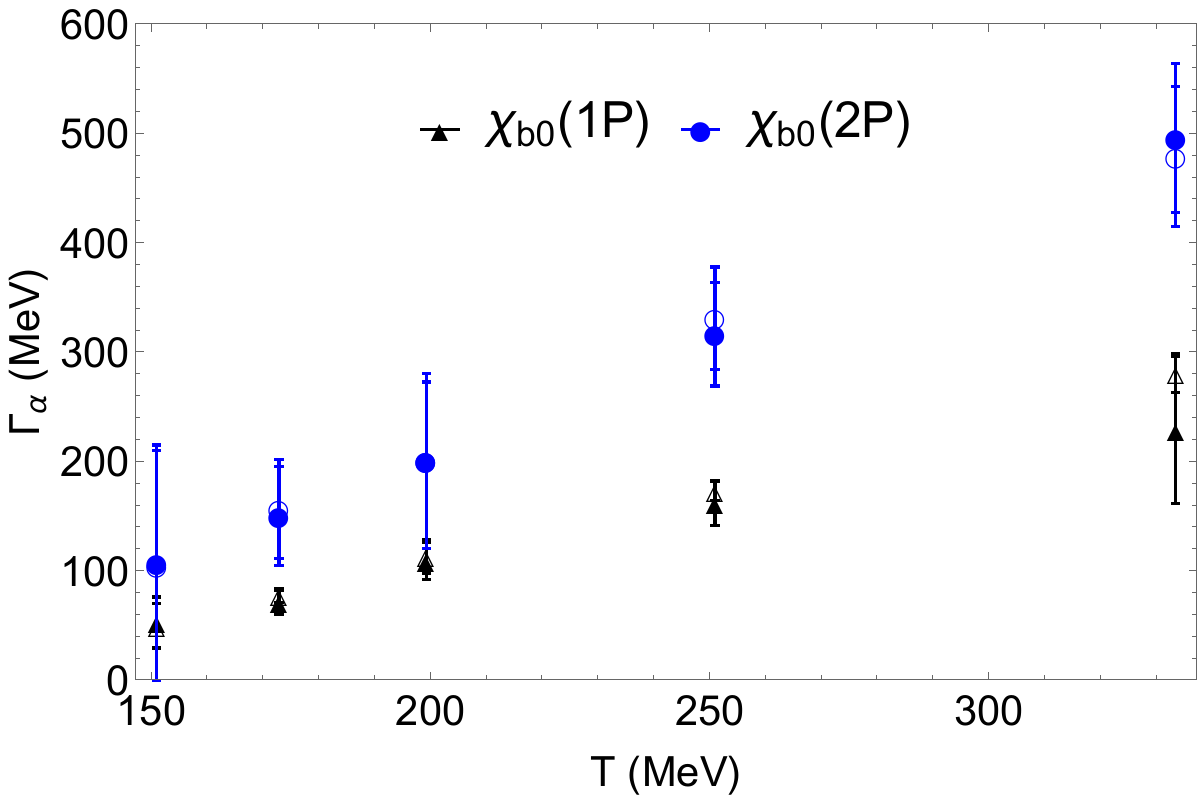}
}
\caption{The mass shift (top) and thermal width (bottom) for ground and excited states of $\Upsilon$ (left) and $\chi_{b0}$ (right) taken from \cite{Larsen:2019zqv}. }
\label{fig:massshift-width}
\end{figure*}

In \cite{Larsen:2019zqv} this study was extended up to 3S and 2P radial excited states for bottomonium. Subtracting off the temperature-independent contribution $C_{\alpha}^{\mathrm{high}}(\tau)$ that can be extracted from the vacuum correlators using a single-exponential model, one is able to filter out a continuum-subtracted correlator $C_{\alpha}^{\mathrm{sub}}(\tau, T)$ that gives contribution from bound state region or below \cite{Larsen:2019bwy} 
\begin{equation}
C_{\alpha}^{\mathrm{sub}}(\tau, T) = C_{\alpha}(\tau, T) - C_{\alpha}^{\mathrm{high}}(\tau).
\end{equation}
Such correlator can be fit using a simple theoretically-motivated Ansatz of the in-medium spectral function 
\begin{equation}
\rho_{\alpha}^{\mathrm{med}}(\omega, T) = A_{\alpha}^{\mathrm{cut}}(T)\delta \left( (\omega- \omega_{\alpha}^{\mathrm{cut}}(T)\right) + A_{\alpha}(T)\exp\left(-\frac{|\omega-M_\alpha(T)|^2}{2\Gamma^2_{\alpha}(T)}\right),
\end{equation}
where the first term provides a simple parametrization for the low frequency tail of the spectral function \cite{Larsen:2019bwy}. From this fit the peak width $\Gamma_{\alpha}(T)$ and peak location $M_\alpha(T)$ of the bound state can be obtained for different states and temperatures. The results are collected in Fig.\ref{fig:massshift-width}. It can be seen that, compared to the mass at zero temperature, the mass at high temperature remains unchanged for all the states studied. Unlike the mass shift, the thermal width at above $T_c$, however, is non-vanishing.  The width increases with temperature for all states. Furthermore, one can see a hierarchical increasing pattern, namely the excited states which have smaller binding energy and larger size, receive larger thermal broadening.

\section{Complex static quark-antiquark potential}

The static quark potential was previously believed to be  real valued, similar as the free energy of quark anti-quark pair. In the modern view point, where the $Q\bar{Q}$ in a thermal bath is treated as an open quantum system, the potential is found to develop an imaginary part 
\cite{Laine:2006ns,Brambilla:2008cx}. In the leading order 
hard thermal loop (HTL) resumed perturbation calculation one gets \cite{Laine:2006ns}
\begin{equation}
\begin{split}
\lim_{t\rightarrow \infty} V_>(t,r) &=-\frac{g^2C_F}{4\pi}\left[ m_D+\frac{\exp(-m_Dr)}{r}\right]-\frac{ig^2TC_F}{4\pi}\phi(m_Dr),\\
\phi(x)&=2\int_0^\infty \frac{dz\ z}{(z^2+1)^2} \left(1-\frac{\sin(zx)}{zx}\right). 
\end{split}
\end{equation}
Here $m_D$ is the Debye mass.
This calculation is valid for distances $r \simeq 1/m_D$.
We see that the real part of the potential in this case is 
screened and is equal to the singlet free energy in Coulomb gauge.
However, in more general case the real part of the potential
is not screened, though it may receive thermal corrections, see
Ref. \cite{Brambilla:2008cx}. 

The existence of both the real and imaginary parts of the potential stimulates a dynamical picture of understanding the quarkonium melting: the dissociation of $Q\bar{Q}$ is due to Landau damping and singlet-to-octet transitions, replacing Matsui and Satz's original static screening picture \cite{Matsui:1986dk}. Unfortunately, these weak-coupling treatments fail to apply for realistic quark masses and temperature.
Thus a non-perturbative determination of the potential is demanded.

The non-perturbative calculation of the potential relies on the lattice formulation proposed by Rothkophf, Hatsuda and Sasaki \cite{Rothkopf:2011db}, which treats the static quarkonium correlator as Wilson loop or Coulomb-gauged thermal Wilson line correlators in heavy quark mass limit
\begin{equation}
 \langle (\bar{Q}Q)(\bar{Q}Q)^\dagger\rangle \stackrel{M_Q \rightarrow \infty}{=} W_\square(r,t). 
\end{equation}
The real-time evolution of the Wilson loop ignoring non-potential effects can be expressed in an in-medium Schrödinger equation
\begin{equation}
 i\partial_t W_\square(r,t) \simeq V(r)W_\square(r,t)
\end{equation}
whose solution takes the form 
\begin{equation}
V_{\square}(r,t) = \frac{i\partial_t W_\square(r,t)}{W_\square(r,t)}
 =\frac{\int d\omega \, \omega\, e^{-i\omega t} \rho_\square(r,\omega)}
 {\int d\omega\, e^{-i\omega t} \rho_\square(r,\omega)}
\label{Eq:PropPot}
\end{equation}
in the spectral representation that defines a non-perturbative in-medium potential
\begin{equation}
 W_\square(r,t) = \int_{-\infty}^{+\infty}
  d\omega e^{-i\omega t} \rho_\square(r,\omega). 
\end{equation}
Performing an analytic continuation of above equation to imaginary time we
obtain the connection of the Euclidean thermal Wilson-loop to the spectral function through the Laplace transform
\begin{equation}
 W^{\mathrm{E}}_\square(r,\tau) 
= \int d\omega e^{-\omega \tau}  \rho_\square(r, \omega).
\end{equation}
The spectral function contains a peak structure. Subtracting out the remnant high frequency vacuum contributions yields the dominant resonance part, whose peak position corresponds to the real part of the potential, while the peak width corresponds to the imaginary part. The real part determines the binding energy of the quarkonium, while the imaginary part gives its thermal width, characterizing the thermal effects on the dilepton production from bottomonium. Using the lattice data on Wilson loops or Wilson line
correlators and modeling of the corresponding
spectral function gives the complex potential.

\begin{figure*}
\centerline{
\includegraphics[width=0.5\textwidth]{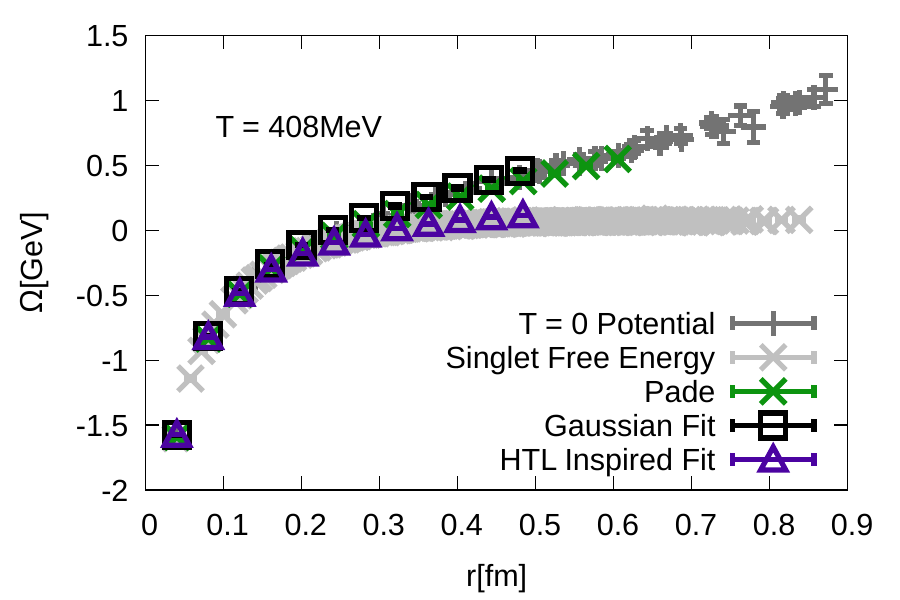}
\includegraphics[width=0.5\textwidth]{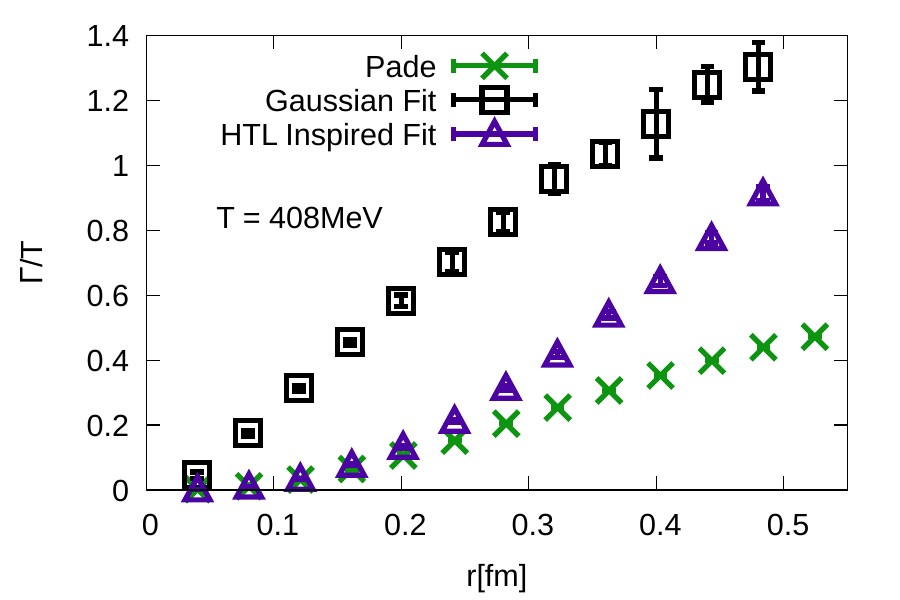}
}
\centerline{
\includegraphics[width=0.5\textwidth]{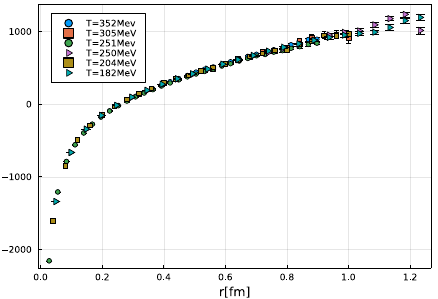}
\includegraphics[width=0.5\textwidth]{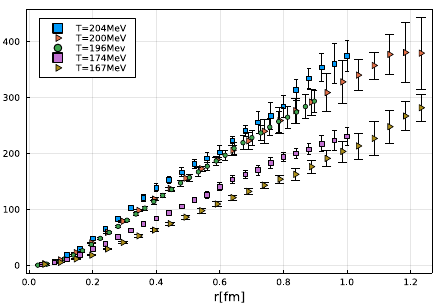}
}
\caption{The real (left) and imaginary (right) part of the complex potential taken from \cite{Bala:2021fkm} (top) and \cite{Bazavov:2023dci} (bottom). }
\label{fig:potential}
\end{figure*}

The lattice computations of complex potential can already be done using realistic setup after more than a decade of development \cite{Rothkopf:2011db,Burnier:2012az,Burnier:2013fca,Burnier:2013nla,Burnier:2015tda,Bala:2021fkm,Bazavov:2023dci}. Various models, including Gaussian and Lorentzian, have been used in the spectral analysis. Pade fit, HTL-inspired fit and Bayesian BR analysis \cite{Burnier:2013nla} are also explored. Results from a thorough analysis using different spectral reconstruction techniques are shown in the top panels of Fig.\ref{fig:potential}, which utilized Pade fit, HTL-inspired fit and Gaussian fit on realistic (2+1)-flavor lattice data. Bayesian BR analysis was also used but failed to provide reliable results at temperature above crossover. This calculation found increasing thermal width with spatial distance $r$ but whether the real part is screened is not conclusive: Pade and Gaussian fit suggests no screening while HTL-inspired fit does.

\begin{figure*}
\centerline{
\includegraphics[width=0.5\textwidth]{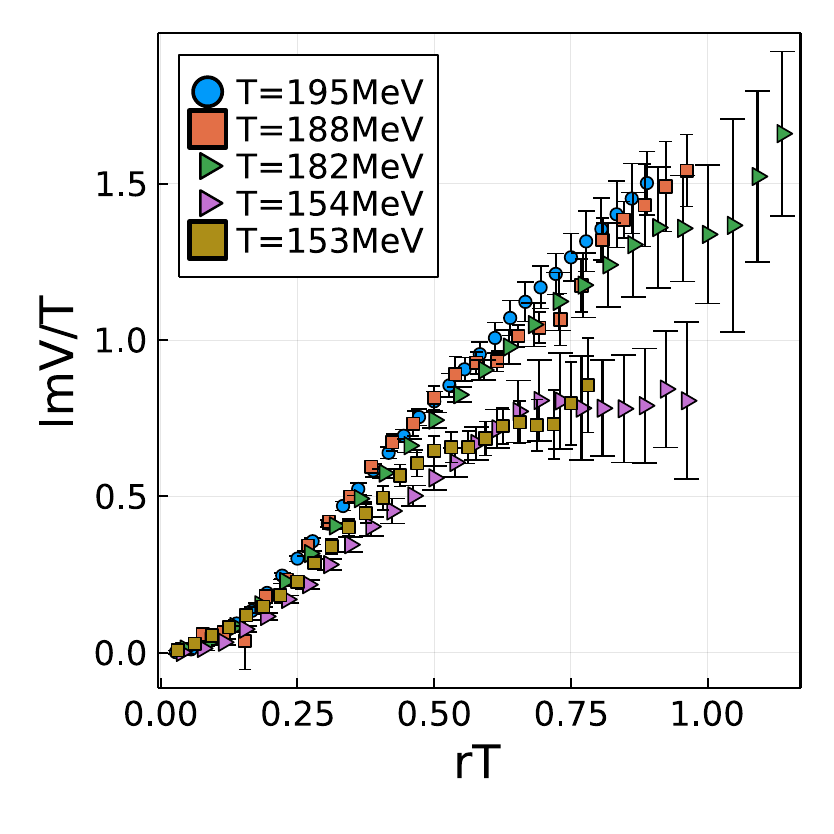}
\includegraphics[width=0.5\textwidth]{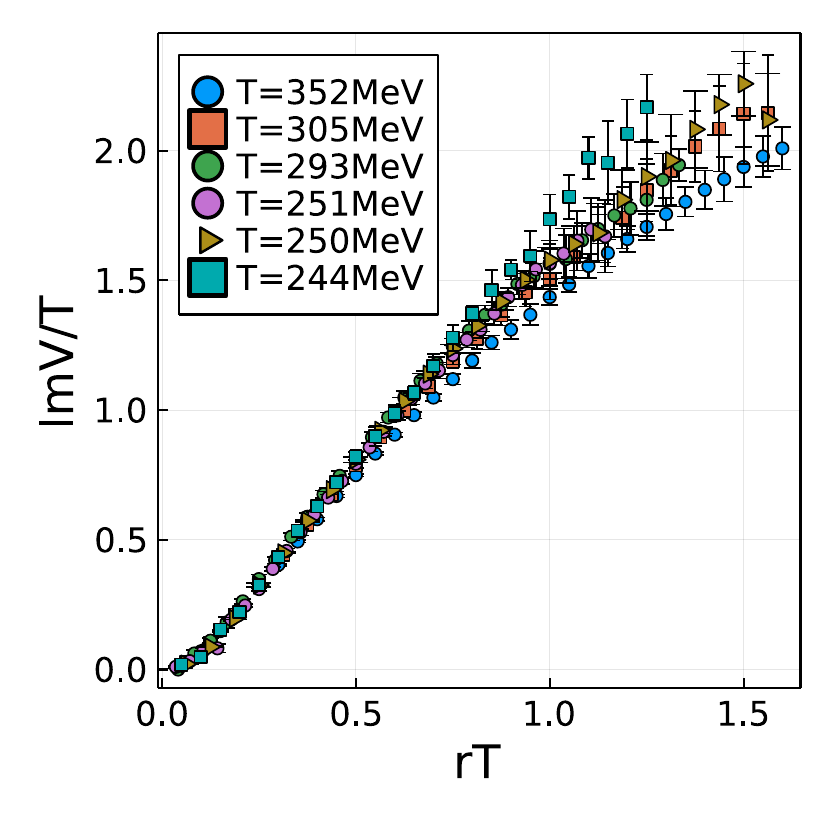}
}
\caption{The imaginary potential scaled by temperature as a function of $rT$ for relatively low temperatures (left) and high temperatures (right) taken from \cite{Bazavov:2023dci}.}
\label{fig:potential-V}
\end{figure*}

The state-of-the-art lattice calculation is carried out by HotQCD collaboration on (2+1)-flavor HISQ configurations at physical pion mass in the temperature range 153 MeV $\leq T\leq$ 352 MeV \cite{Bazavov:2023dci}. This calculation employs a so-called cut Lorentzian Ansatz for the dominant peak structure
\begin{align}
\rho_r^{\text{peak}}(\omega,T) = \frac{1}{\pi} \frac{A_r(T) \Gamma(\omega,r,T)}{[\omega-{\rm Re}V(r,T)]^2+\Gamma^2(\omega,r,T)},
\end{align}
which treats the real part the potential $\mathrm{Re}V(r,T)$ as a fit parameter and whose second cumulant can be interpreted as the imaginary part of the potential $\mathrm{Im}V(r,T)$. Regions beyond the peak vicinity \mbox{$|\omega-{\rm Re} V(r,T)| \gtrsim \Gamma(r,T)$} are set to zero. This is consistent with the T-matrix approach \cite{Liu:2017qah} observation that the spectral function dies off soon when away from the peak. The results are shown in the bottom panels of Fig.\ref{fig:potential}. This calculation found increasing thermal width with $r$, consistent with the findings of \cite{Bala:2021fkm}, and it increases more for higher temperatures. This suggests stronger thermal broadening of static quark-antiquark pair for higher temperature, corroborating the increasing thermal width seen for $\Upsilon$ shown in Fig.\ref{fig:massshift-width}. In Fig.\ref{fig:potential-V} the imaginary potential scaled by temperature is shown in $rT$. It can be seen that for $T>180$ MeV (the pseudo $T_c$ of this lattice setup) at $rT\approx 1$, the imaginary potential is larger than the temperature, implying a quick damping of force between $Q$ and $\bar{Q}$. In such a short time, the chromo-electric field between $Q\bar{Q}$ has no time to react to the presence of the medium before the force could forge bonds between $Q$ and $\bar{Q}$, leading to the melting of quarkonium. Such a picture has a distinct difference than what Matsui and Satz conjectured.

We note that the ground peak location is not sensitive to the detailed shape if one uses a Gaussian or Lorentzian Ansatz with proper cut. The HTL-inspired model introduces a screening by definition, but is ruled out for that it does not fit the lattice 
data even at very high temperature $T=2$ GeV \cite{Bala:2021fkm}. The absence of color screening explains why there is no mass shift for bottomonium seen in Fig.\ref{fig:massshift-width}.

\section{Heavy quark diffusion}

\begin{figure*}
\centerline{
\includegraphics[width=0.5\textwidth]{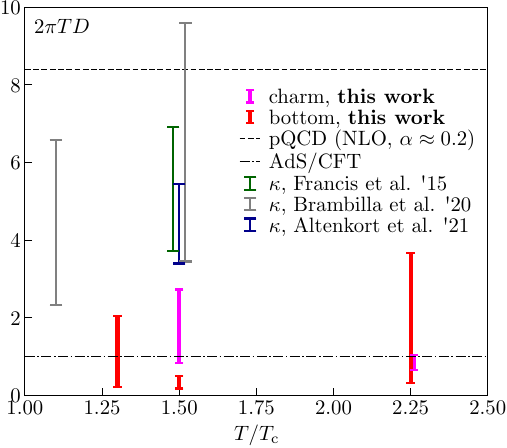}
\includegraphics[width=0.5\textwidth]{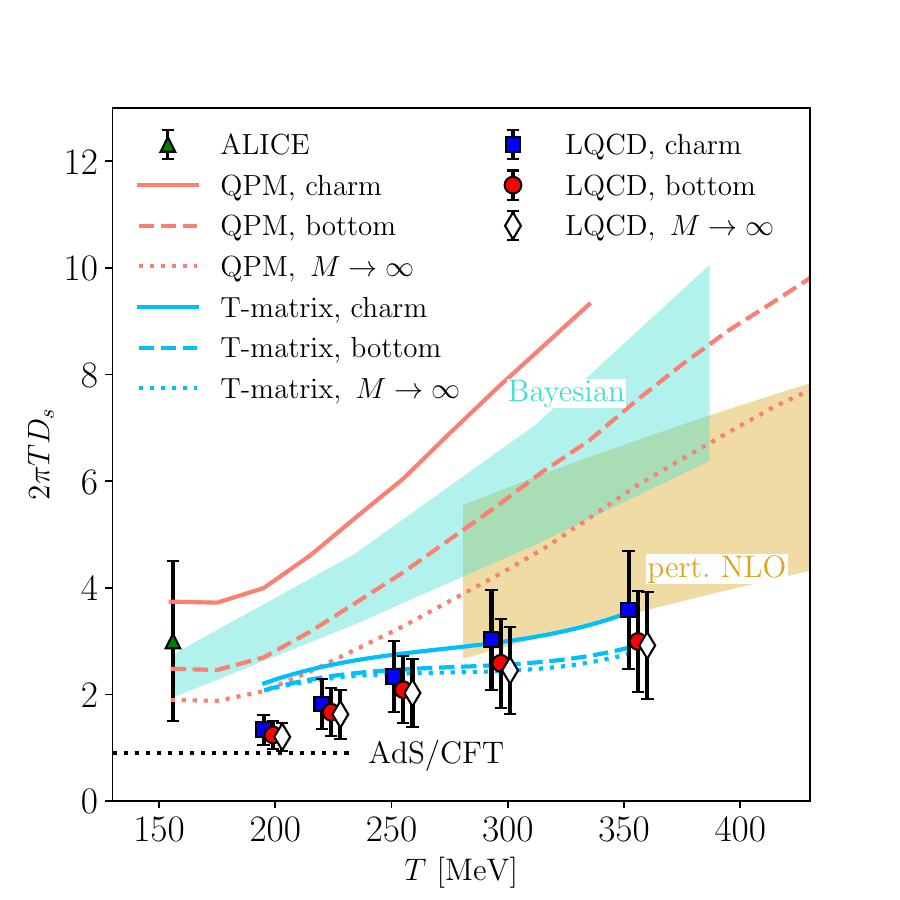}
}
\caption{Left: heavy quark diffusion coefficient calculated in the quenched limit on the lattice (vertical lines) \cite{Francis:2015daa, Altenkort:2020fgs, Brambilla:2020siz} and in AdS/CFT \cite{Casalderrey-Solana:2006fio} and NLO pQCD \cite{Caron-Huot:2007rwy}. Right: (2+1)-flavor lattice QCD results (points) \cite{Altenkort:2023oms,Altenkort:2023eav}, quasi particle model results \cite{Sambataro:2023tlv}, T-matrix calculations \cite{Liu:2016ysz, ZhanduoTang:2023ewm}, phenomenological estimate \cite{Xu:2017obm,ALICE:2021rxa} and Ads/CFT result \cite{Casalderrey-Solana:2006fio}.}
\label{fig:2piTD}
\end{figure*}

Heavy quark diffusion coefficient $D_s$ tells how fast the heavy quarks get thermalized in the medium after produced in the very early stages of the heavy ion collisions or released from
the bound state. Its determination has a long history of almost
two decades, see e.g. \cite{Petreczky:2005nh} and references therein. 
$D_s$ can be determined from the low frequency part of the spectral function embedded in the vector current-current correlators via Kubo-formula
\begin{equation}
D_s=\frac{1}{3\chi_q}\lim_{\omega\rightarrow 0}\sum_{i=1}^3\frac{\rho^{ii}(\omega)}{\omega}.
\end{equation}
Due to the difficulty of putting relativistic heavy quarks on the lattice, the up-to-date calculation with physical charm and bottom quark mass is still in the quenched limit \cite{Ding:2021ise}. In \cite{Ding:2021ise} $D_s$ is extracted from the transport peak of the spectral function, whose contribution can be obtained by subtracting off the high frequency contribution (modeled by pNRQCD and ultraviolet asymptotics) from the whole correlator. Assuming a  Lorentzian form for the transport peak, $D_s$ 
can be fit with quark mass taken from PDG book. The results are shown in the left panel of Fig.\ref{fig:2piTD} as thick colorful lines, which lie around the AdS/CFT result \cite{Casalderrey-Solana:2006fio}. There are already attempts to extend such calculation to full QCD starting from  pseudo-scalar channel first \cite{Ali:2023kmr}.

The heavy quark diffusion can also be accessed in a simpler way. In the infinite heavy mass limit heavy quark effective theory (HQEF) applies. Combined with Langevin dynamics a momentum diffusion coefficient $\kappa$ can be defined that relates to $D_s$ through \cite{Moore:2004tg}
\begin{equation}
\kappa= \frac{2 T^2}{D_s} \cdot\frac{\langle p^2 \rangle}{3MT}.
\end{equation}
$\kappa$ was first calculated in the strong-coupled $\mathcal{N}=4$ Yang-Mills theory \cite{Casalderrey-Solana:2006fio} and then in perturbation theory at next-to-leading order (NLO) \cite{Caron-Huot:2007rwy}. Soon it was found that it can be calculated non-perturbatively on the lattice \cite{Caron-Huot:2009ncn} by measuring the color-electric field correlators 
\begin{equation}
    G_E(\tau, T) = -\sum_{i=1}^{3} 
    \frac{\left\langle {\rm Re Tr}\left[U(\beta,\tau)E_i(\mathbf{x},\tau)U(\tau,0)E_i(\mathbf{x},0)\right]\right\rangle}{3\left\langle {\rm Re Tr} U(\beta,0)\right\rangle}.
    \label{eq:gblat-e}
\end{equation}
Again the low frequency part of the spectral function encoded in $G_E(\tau, T)$ gives the transport coefficient \cite{Caron-Huot:2009ncn}
\begin{equation}
\label{kappa}
    \kappa_E = \lim_{\omega \rightarrow 0} 2T\frac{\rho_E}{\omega}.
\end{equation}
At large frequency regime it has a rather simple structure $\rho_E(\omega)\propto \omega^3$ at LO and is also known at NLO \cite{Burnier:2010rp}. Various strategies can be used to control the uncertainty of connecting these two regimes. The quenched calculations of $\kappa_E$ include Refs.\cite{Francis:2015daa, Altenkort:2020fgs, Brambilla:2020siz, Banerjee:2022gen, Brambilla:2022xbd}. Results from some of them are shown as thin vertical lines in the left panel of Fig.\ref{fig:2piTD}. We can see that when converted to $2\pi TD_s$ they are larger than those obtained from current-current correlators.

\begin{figure}[h]
\centerline{
\includegraphics[width=0.5\textwidth]{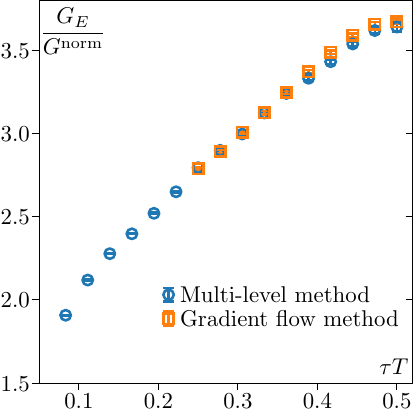}
}
\caption{A comparison of color-electric correlators calculated using multilevel algorithm and gradient flow method in the quenched approximation at 1.5$Tc$ \cite{Altenkort:2020fgs}.}
\label{fig:flow-vs-ml}
\end{figure}

Since physical quarks like charm and bottom have finite mass, the finite mass correction should be estimated for a realistic purpose. In \cite{Bouttefeux:2020ycy} it was shown that the $\kappa_E$ calculated above gives the leading order contribution when expanding the force-force correlator (an intermediate quantity in the derivation of $\kappa$) in $T/M$. 
The finite mass correction $\kappa_B$ can be determined from the color-magnetic field correlator
\begin{equation}
G_B(\tau, T) = \sum_{i=1}^{3} 
    \frac{\left\langle {\rm Re Tr}\left[U(\beta,\tau)B_i(\mathbf{x},\tau)U(\tau,0)B_i(\mathbf{x},0)\right]\right\rangle}{3\left\langle {\rm Re Tr} U(\beta,0)\right\rangle}.
    \label{eq:gblat-b}
\end{equation}
The spectral function of this correlator has the same form as $\rho_E$ at LO and is slightly different at NLO \cite{Banerjee:2022uge}. Quenched calculations of $\kappa_B$ can be found in \cite{Banerjee:2022uge, Brambilla:2022xbd}. The complete momentum diffusion coefficient reads
\begin{equation}
\kappa=\kappa_E+\frac{2}{3} \langle v^2\rangle \kappa_B.
\label{eq:kappa}
\end{equation}
Since $G_E$ and $G_B$ directly involve lattice gauge fields, both correlators suffer strong ultraviolet fluctuations on the lattice. A precise determination  of the correlators
is crucial for a reliable spectral reconstruction. But this requires noise reduction techniques. In the early studies multilevel algorithm \cite{Luscher:2001up} was used, which only applies in quenched approximation without substantial adjustment \cite{Ce:2016idq}. Recently the gradient flow method \cite{Narayanan:2006rf,Luscher:2009eq,Luscher:2010iy,Luscher:2011bx} was introduced in the calculations of transport coefficients \cite{Altenkort:2020fgs,Altenkort:2020axj,BarrosoMancha:2022mbj,Bonanno:2023thi,Itou:2020azb}. The advantage of this method is that it can smear both gauge fields and quark fields continuously, and renormalize them at the same time. In Fig.\ref{fig:flow-vs-ml} we show a comparison of the color-electric correlators calculated from both multilevel and gradient flow in the quenched approximation at 1.5$T_c$ \cite{Altenkort:2020fgs}. A 
very good agreement of the two methods can be seen. 

With the help of gradient flow, the calculation of $\kappa_E$ was extended to full QCD for the first time in \cite{Altenkort:2023oms}. This calculation was performed using (2+1)-flavor light dynamical quarks corresponding to a pion mass of around 320 MeV in the temperature range $195\ \mathrm{MeV}<T<352\ \mathrm{MeV}$. The results are shown as open diamonds in the right panel of Fig.\ref{fig:2piTD}. It can be seen that, compared to the quenched results shown in the left panel, the full QCD results are much smaller if plotted in $T/T_c$ (in this study $T_c=180$ MeV). It should be pointed out that when plotted in absolute temperature, quenched results and full QCD results follow a smooth increasing pattern with temperature.

The calculation of the mass correction involves a non-trivial anomalous dimension at NLO renormalization \cite{Guazzini:2007bu,Banerjee:2022uge}.
This leads to a problematic zero flow-time limit extrapolation of $G_B$.
As a compromise, in \cite{Brambilla:2022xbd} $\kappa_B$ was calculated at finite flow time and then extrapolated to zero flow time.
In Ref. \cite{Banerjee:2022uge} multilevel algorithm and non-perturbative
renormalization of $G_B$ was used.

In \cite{Altenkort:2023eav}, a three-step matching procedure was proposed that solves this issue at the correlator level and converts the flowed one to its physical value via a factor $Z_{\mathrm{match}}$
\begin{equation}
G_B^{\mathrm{phys}}(\tau,T) = \lim_{\tau_{\mathrm{F}} \rightarrow 0} 
Z_{\mathrm{match}}({\bar{\mu}}_T, {\bar{\mu}}_{\tau_{\mathrm{F}}},\mu_{\mathrm{F}}) \, G_B(\tau,T,\tau_{\mathrm{F}}).
\label{eq:match-to-physical}
\end{equation}
With proper choices for the intermediate scales ${\bar{\mu}}_T$ and ${\bar{\mu}}_{\tau_{\mathrm{F}}}$, a reasonable control on the uncertainty in the matching can be achieved. Based on this, a full QCD determination of $\kappa_B$ is obtained in \cite{Altenkort:2023eav} using the same lattice setup as in \cite{Altenkort:2023oms}. $\kappa_B$ turns out to be of similar size as $\kappa_E$ at the same temperature (not shown here but can be found in \cite{Altenkort:2023eav}). Together with $\langle v^2\rangle$ and $\langle p^2\rangle$ calculated in the quasi particle model, the heavy quark diffusion coefficient for physical charm and bottom quark is obtained. The results are shown as blue squares and red circles for charm and bottom in the right panel of Fig.\ref{fig:2piTD}. We can see that even though $\kappa_E$ and $\kappa_B$ are of similar size, the increment from $\kappa_B$ scaled by 2/3$\langle v^2\rangle$ is compromised by the increment of $\langle p^2\rangle/(3MT)$, rendering the final $2\pi TD_s$ remarkably independent
of the heavy quark mass.

The lattice results have similar increasing pattern in temperature and decreasing pattern in quark mass as the quasi particle model results \cite{Sambataro:2023tlv} and T-matrix calculations \cite{Liu:2016ysz, ZhanduoTang:2023ewm}. At low temperature around $T_c$ lattice results are consistent with AdS/CFT calculation \cite{Casalderrey-Solana:2006fio} and the ALICE collaboration's phenomenological estimate \cite{Xu:2017obm,ALICE:2021rxa} within error. At high temperature they stretch into the T-matrix approach results and NLO perturbative estimate \cite{Caron-Huot:2007rwy}. The small value of $2\pi TD_s$ at around crossover temperature suggests that the QGP comes to an equilibrium extremely rapidly after creation. The very short mean free path of the heavy quark hints that QGP is a nearly-perfect fluid.

\section{Summary}

In these proceedings, we have reviewed the recent progress 
on the lattice QCD calculations of the heavy flavor probes. Three topics including in-medium quarkonium properties, complex in-medium inter-quark potential and heavy quark diffusion have been discussed. Former two are related to each other involving the melting mechanism of quarkonium. The last one addresses the transport phenomena in QGP. All these studies rely on spectral reconstruction based on Eq.(\ref{eq:corr-spec}). 

To access the in-medium quarkonium spectral function, quarkonium current-current correlators need to be calculated on the lattice. The technique of extended source operator and NRQCD are shown to be able to gain us better resolution of the properties of quarkonium states. In this setup the calculations can already be done using (2+1)-flavor quark action at physical pion mass \cite{Larsen:2019bwy,Larsen:2019zqv}. Simple modeling shows that ground and excited states of $\Upsilon$ and $\chi_{b0}$ do not have mass shift up to 333 MeV. The thermal width, on the other hand, increases with temperature, and it increases faster for higher excited states. Studies along the tradition way, namely using relativistic heavy quarks, have also become available for full QCD, first in the pseudo-scalar channel \cite{Ali:2023kmr} and the vector channel will come soon. Access to the full frequency space of the spectral function allows for the determination of both the diffusion coefficient and the thermal modifications  of quarkonium bound states. The up-to-date calculations show that no resonance peaks for $J/\Psi$ are needed at $T\geq 1.1T_c$, while for $\Upsilon$ a resonance peak is still needed up to $1.5Tc$, consistent with previous quenched results. 

Using the lattice formulation proposed in \cite{Rothkopf:2011db}, the inter-quark potential can be extracted from Wilson loop or Coulomb-gauged thermal Wilson line correlators, which can be non-perturbatively calculated on the lattice. In the most recent calculation \cite{Bazavov:2023dci} using (2+1)-flavor HISQ configurations at physical pion mass and at temperature 153 MeV $\leq T\leq$ 352 MeV, a detailed discussion on the applicability and consistency of various model spectral functions was given. This calculation concludes that there is no screening for the real part of the potential in the entire temperature range up to $r\approx 1.0$ fm. The unscreening is not sensitive to the detailed shape of the ground peak. Using a Gaussian Ansatz or a Lorentzian Ansatz with proper cut for the spectral function
leads to the same result. A HTL-inspired Ansatz introduces intrinsic screening but does not describe the lattice data even at very high temperature scale up to 2 GeV. The imaginary part increases with $r$ and the temperature. The new results corroborate the observation of zero mass shift and increasing thermal width seen above from the quarkonium correlators. They also give rise to a new dynamical melting picture of quarkonium that substitutes Matsui and Satz's original screening conjecture.

The heavy quark diffusion coefficient can be extracted from either the vector current-current correlators or from the color-electric field correlators and color-magnetic field correlators. The former case is still limited to quenched approximation due to the complexity of having relativistic heavy quarks on the lattice but a first attempt can be expected to arrive soon. In the latter case full QCD results at unphysical pion mass are already available for charm and bottom quark \cite{Altenkort:2023oms,Altenkort:2023eav} with the help of gradient flow method. It was shown for the first time that the full QCD $2\pi TD_s$ has very weak dependence on quark mass and slowly increases with temperature. Its small value at around crossover temperature suggests a fast thermalization of heavy quarks and verifies that QGP is a nearly-perfect fluid. This calculation can be further extended to physical pion mass with which lower temperature down to and/or below physical crossover temperature $T=153$ MeV can be reached.

\section{Acknowledgments}
We thank P. Petreczky for discussions, careful reading and comments on the manuscript. HTS’s research is supported by The U.S. Department of Energy, Office of Science, Office of Nuclear Physics through Contract No.~DE-SC0012704, and within the frameworks of Scientific Discovery through Advanced Computing (SciDAC) award \textit{Fundamental Nuclear Physics at the Exascale and Beyond} and the Topical Collaboration in Nuclear Theory \textit{Heavy-Flavor Theory (HEFTY) for QCD Matter}.

\bibliographystyle{apsrev4-1}
\bibliography{references}

\end{document}